\documentstyle[epsfig]{aipproc}

\begin{document}

\def\wwtt{{\mbox {\,$W^+W^-t{\bar t} \;$}\,}}
\def\ra{\rightarrow}
\def\bea{\begin{eqnarray}}
\def\eea{\end{eqnarray}}

\noindent
ANL-HEP-CP-01-006 \hfill hep-ph/0101253

\title{The Search for Anomalous $W^+ W^- t \bar{t}$\\[0.2cm]
Couplings at the LC\thanks{Talk 
contributed to the Linear Collider Workshop 2000
(LCWS2000), 24-28 October 2000, Fermilab, Batavia, Illinois, USA.}}

\author{Francisco Larios$^1$, Tim M.P. Tait$^2$, and C.--P. 
Yuan$^3$}
\address{
$^1$ Departmento de F\'{\i}sica Aplicada, CINVESTAV-M\'erida\\
A.P. 73,  97310 M\'erida, Yucat\'an,  M\'exico\\
\vspace*{0.4cm}
$^2$ HEP Division, Argonne National Lab\\
9700 South Cass Avenue, Argonne, IL 60439
USA\\
\vspace*{0.4cm}
$^3$ Dept. of Physics and Astronomy,\\
Michigan State University\\
East Lansing, MI 48824 USA\\
}

\maketitle

\begin{abstract}
We study production of $t\bar t$ via $W^{+} W^{-}$-fusion,
including the relevant backgrounds at the proposed
Linear Collider (an $e^+ \, e^-$ collider with
$\sqrt{S}\,=\,1.5$ TeV) in the context of a {\it Higssless}
Standard Model (HSM), i.e. a nonlinear SU(2$)_L \times$ U(1$)_Y$ 
chiral Lagrangian, including dimension five $W^+ W^- t \bar{t}$ 
interactions.  Deviation from the HSM predictions can be used to 
constrain the coefficients of these operators to an order of $10^{-1}$ 
(divided by the cut-off scale $\Lambda = 3.1 \;$TeV) with a $95\%$ C.L..
\end{abstract}

One of the deep mysteries with which we are confronted today 
in particle physics is the mechanism of the electroweak 
symmetry breaking (EWSB).  While the fundamental Higgs mechanism 
postulated by the Standard Model (SM) currently provides an 
excellent description of all experimental data, the Higgs itself 
has defied discovery.  This leads us to explore other probes of the
EWSB sector, such as deviations in the couplings of known particles
from SM predictions, induced by the underlying EWSB physics.

The top quark, as the only quark whose mass is of the same order as the
EWSB scale, $v \sim 246$ GeV, provides a natural laboratory in which to
search for these nonstandard effects.  In fact, the large top mass may
be a clue that top plays an essential role in the EWSB, and a detailed
study of its properties may prove to be the only way to unravel the
true nature of the EWSB dynamics.  It is important that these interactions
be studied exhaustively and model-independently, since without a deep
understanding of the underlying physics it is impossible to know
what to expect.

In this talk we address how one
can use the process $W^+ W^- \rightarrow t \bar{t}$ 
at a high energy $e^+ e^-$ collider (which we assume to have 
center-of-mass energy $\sqrt{S} = 1.5$ TeV with a data sample of 
200 ${\rm fb}^{-1}$)
to probe several parameters
of a generic effective theory in which the electroweak symmetry is
broken, but no assumptions about how the breaking occurs 
are imposed.  We show that non-standard interactions of 
${\cal O}(10^{-1})$ are tested by this process
\cite{wwtt}.  Our model-independent results, in the context of a
nonlinear realization of the electroweak (EW) gauge symmetry \cite{nonlinear}
are complimentary to studies based on specific models \cite{wwtt2},
upon parameterizations with linear realizations of the 
EW symmetry \cite{wwtt3},
studies which include generic scalar bosons \cite{wwtt4},
as well as to studies at hadron colliders based on single top production
\cite{singlet}.
If there is new physics
associated with the EWSB present in this channel, it could be discovered
in this way, and we would know what some of the properties an 
underlying theory  would have to possess in order to explain the 
observed deviation.  Similarly, if no effect is discovered, we effectively
constrain any theory of new physics which contributes to this process.

Our dimension five interactions are part of the electroweak
chiral Lagrangian, which realizes the
EW gauge symmetry nonlinearly.  This parameterization is most appropriate
for a theory in which the underlying EWSB dynamics is strong, 
and there is effectively no Higgs boson at low energies,
and thus our setting can
be thought of as the {\em Higgsless} Standard Model (HSM).  
As was studied in Ref.
\cite{top5}, there are fourteen dimension five operators which contribute to
$\wwtt$ scattering in this context, 
however only seven of them contribute to the leading
behavior in $E$, the energy of the $t \bar{t}$ system.  These seven operators
may be further classified by their Lorentz structure as either scalar
operators, contributing only to S-wave $W^+_L W^-_L \ra  t \bar{t}$ 
scattering, as tensor operators contributing to P-wave scattering, and
one operator which contains both structures.  Thus, without loss of generality,
we restrict our attention to one dimension 5 operator of each type,
contributing at leading order in $E$,
\bea
\label{dim5}
{\cal L}^5_{eff}&=&
{\frac {a_1}{\Lambda}}
\bar t  t {W}_{\mu}^{+} {W}^{- \mu } +
{\frac {a_2}{\Lambda}}
i \bar t  {\sigma}^{\mu\nu} t {W}_{\mu}^{+} 
{W}^{-}_{\nu }\; ,
\eea
where $a_1$ and $a_2$ parameterize the strength of the S-wave and P-wave
operators, respectively, and are expected from naive dimensional analysis
to be order 1 when the cut-off of the theory is
$\Lambda = 4 \pi v = 3.1$ TeV \cite{georgi}.  In our discussion below,
we set $\Lambda = 3.1$ TeV and discuss how well one may constrain
$a_1$ and $a_2$ for this assumed value of $\Lambda$.

\begin{figure}
\epsfysize=1.0in
\centerline{\epsfbox{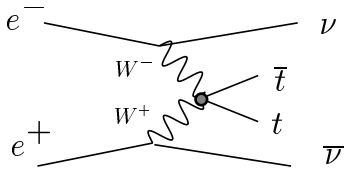}}
\caption{Diagrams for $e^+ e^- \ra \nu_e \bar{\nu}_e t \bar{t}$ through
anomalous \wwtt couplings.}
\label{fig:feyn}
\end{figure}

We perform detailed Monte Carlo simulations of the process 
$e^+ e^- \ra \nu_e \bar{\nu}_e t \bar{t}$, including the exact $2 \ra 4$
tree level matrix elements.  The HSM contains 19 Feynman
diagrams for this process, and our new physics that we wish to probe
contributes up to two extra diagrams, as shown in Fig.~\ref{fig:feyn}, one
for each operator under study.  Assuming that the new physics contributions
are small compared to the HSM rates, our dominant signal
will arise from the interference between the HSM graphs, and ones containing
the new physics operators in Eq.~\ref{dim5}.

Our
experimental signature is taken to be a $t \bar{t}$ pair, and missing
transverse energy ($E_T$) from the neutrinos in the final state.  
We also consider
the dominant background to this process, $e^+ e^- \ra t \bar{t} \gamma$,
where the photon is missed because it falls outside of the detector coverage
(assumed to be 0.15 rad about the beam axis),
thus faking missing $E_T$.  We impose the restriction that the $t$ and
$\bar{t}$ each have transverse momentum $p_T \geq 20$ GeV and rapidity
$|y| \leq 2$ in order to be observable.  In addition, we find that we
can suppress the fake background by requiring the $p_T$ of the $t \bar{t}$
pair to also be larger than 20 GeV, because the fake rate produces
photons which tend to be collinear with the incoming electrons.  After 
imposing these cuts, we are left with cross sections of 4 fb (assuming
$a_1 = a_2 = 0$) for
$e^+ e^- \ra t \bar{t} \nu \bar{\nu}$ and 
1.8 fb for $e^+ e^- \ra t \bar{t} \gamma$.  The fake background may be
further reduced, by polarizing the $e^-$ and/or $e^+$ beam, because in the
signal $W^+ W^- \ra t \bar{t}$ process the $W$ bosons only couple to
left- (right-handed) fermions (anti-fermions), whereas the fake process
is dominated by initial state radiation which couples to both polarizations.

\begin{figure}
\epsfysize=3.0in
\centerline{\epsfbox{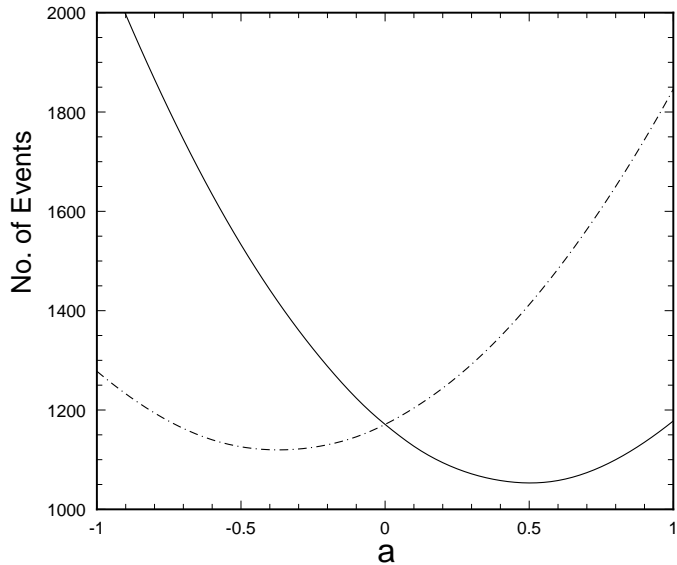}}
\caption{Number of $W^+ W^-$ fusion $t \bar t$ pairs as a function
of the anomalous couplings $a_{1}$ (solid line) and
$a_{2}$ (dash-dot line).  The point $a_1 = a_2 = 0$
corresponds to the prediction of the {\it Higgsless} SM.}
\label{fig:events}
\end{figure}

We now allow $a_1$ and $a_2$ to be non-zero, and derive 95\% 
confidence level (C.L.) constraints on these parameters.  The simplest
thing one can do is to count the number of $W^+ W^- \ra t \bar{t}$
events.  The dependence on $a_1$ and $a_2$ is shown in 
Fig.~\ref{fig:events}.  Note that because the primary influence of the new
physics operators is through an interference effect, the event rate can
actually be smaller than the HSM prediction.
Requiring that no 95\% 
C.L. deviation is observed restricts $-0.13 \leq a_1 \leq 0.18$ and
$-0.9 \leq a_2 \leq 0.2$.

One may improve these constraints by noting that the new physics operators
result in modified kinematic distributions for the produced top quarks.
In particular, we have found that both the rapidity of the $t \bar{t}$
system, $y_{t \bar{t}}$, and the rapidity gap between the 
$t$ and $\bar{t}$,  $y_t - y_{\bar{t}}$, are sensitive
to the presence of the anomalous $\wwtt$ operators.  In particular, the
tensor operator ($a_2$) tends to shift the $t$ and $\bar{t}$ rapidities in the
opposite directions from the HSM predictions, and thus the rapidity gap
distribution is especially effective at 
revealing the presence of such a P-wave new physics effect.  We quantify
the deviation of a given measured distribution compared to the HSM prediction
by the $\chi^2$ deviation,
\bea
\chi^2 \,=\, \Sigma^K_{j = 1}
\frac{(N^{A}_j - N^{HSM}_j)^2}{N^{HSM}_j} ,
\eea
and find that for an integrated luminosity of $L = 200 \, {\rm fb}^{-1}$,
it is optimal to use 3 equally sized bins in the region
$-0.6 \leq y_{t \bar t} \leq 0.6$, and 4 bins for
the $y_t - y_{\bar t} \;$ analysis (in the region
$0 \leq y_t - y_{\bar t} \leq 2$).  Applying the $\chi^2$ analysis improves
the constraints to $-0.10 \leq a_1 \leq 0.12$ from $\chi^2(y_{t\bar t})$
and $-0.3 \leq a_2 \leq 0.2$ from $\chi^2(y_{t}-y_{\bar t})$.  The fact that
the two operators are more evident in two different distributions could also
be used to disentangle one from the other, should a deviation in the total
rate be observed.  We have also examined the polarizations of the $t$ and
$\bar{t}$ and found that some improvement in sensitivity is possible if these 
polarizations are also measured \cite{wwtt}.

In conclusion, we have found that in models in which the EWSB is a result
of underlying strong dynamics, the process $W^+ W^- \ra t \bar{t}$ at a high
energy linear collider is a powerful test of several operators which
may be present in the low energy effective theory, and may provide
useful information in divining the true nature of the EWSB.

This work was supported in part by 
Conacyt and the NSF under grant PHY-980256-4.
Research at Argonne National Lab is funded by the DOE under 
contract W-31-109-ENG-38.  


\end{document}